\documentclass[showpacs]{revtex4}
\usepackage{graphicx}
\usepackage{dcolumn}
\usepackage{amsmath}

\begin{document}

\title{Geometrical aspects of isoscaling}
\author{A. D\'avila, C. Escudero,  J. A.  L\'opez}
\affiliation{Department of Physics, University of Texas at El Paso, El Paso, Texas
79968, U.S.A.}
\author{ C. O. Dorso}
\affiliation{Departamento de F\'isica, FCEN, Universidad de Buenos Aires, N\'u\~nez, Argentina}
\date{\today}

\begin{abstract}
The property of isoscaling in nuclear fragmentation is studied using
a simple bond percolation model with ``isospin'' added as an extra
degree of freedom. It is shown analytically, first, that isoscaling
is expected to exist in such a simple model with the only assumption
of fair sampling with homogeneous probabilities. Second, numerical
percolations of hundreds of thousands of grids of different sizes
and with different $N$ to $Z$ ratios confirm this prediction with
remarkable agreement. It is thus concluded that isoscaling emerges
from the simple assumption of fair sampling with homogeneous
probabilities, a requirement which, if put in the nomenclature of
the minimum information theory, translates simply into the existence
of equiprobable configurations in maximum entropy states.
\end{abstract}

\maketitle


\section{Introduction}

The experimental determination of isoscaling~\cite%
{xu,johnston,laforest,tsang,tsang2001} has prompted a vigorous study of its
origins and its implications on the equation of state of asymmetric nuclear
matter. Isoscaling is the property that fragment yields of similar, but
isotopically different, reactions depend exponentially on the neutron ($N$)
and proton ($Z$) numbers through $R_{21}(N,Z)= {Y_2(N, Z)}/{Y_1(N, Z)}
\approx Exp[\alpha N + \beta Z]$, where $\alpha$ and $\beta$ are fitting
parameters.

In the past, this power law expression for $R_{21}$ has been linked, under
diverse approximations, to primary yields produced by disassembling infinite
equilibrated systems in microcanonical and grand canonical ensembles~\cite%
{tsang,souza}, as well as in canonical ensembles~\cite{das}, and it
has also been observed in the framework of the grand-canonical limit
of the statistical multifragmentation model~\cite{botvina}, in the
expanding-emitting source model~\cite{tsang}, and in the
antisymmetrized molecular dynamics model~\cite{ono}. Furthermore,
under these approximations, the isoscaling parameters $\alpha $ and
$\beta $ have been
found to be related to the symmetry term of the nuclear binding energy~\cite%
{tsang2001,botvina}, to the level of isospin equilibration~\cite{souliotis},
and to the values of transport coefficients~\cite{veselsky}.

More recently, however, it has been determined through the use of
molecular dynamics~\cite{dorso2005} that isoscaling can exist in
purely classical systems, and that it can be created in systems
fully out of equilibrium. It was also found, among other things,
that $R_{21}$ can maintain the power-law behavior even when it
contains yield contributions generated at different times and
corresponding to diverse thermodynamic conditions.

The implications of these findings are many and very important.
Isoscaling is not a quantum process; $\beta$-decay, Pauli's
exclusion principle and its implications for isospin selection
cannot be possible causes of isoscaling. Isoscaling exists in finite
systems out of equilibrium; expanding systems with rapidly varying
temperatures and chemical potentials obey isoscaling. The isoscaling
ratio, $R_{21}$, contains contributions from different times of the
reaction; its final value does not necessarily correspond to the
thermodynamic conditions of a period of the reaction. This, of
course, if not invalidates the scenarios and conclusions presented
by previous studies, at least casts a shadow of doubt on them and,
especially, on the assumed physical meanings of the isoscaling
parameters $\alpha$ and $ \beta$.

In view of the present situation, the question that needs to be
addressed remains the same as in our previous
study~\cite{dorso2005}, what produces isoscaling? Answering this
question is now, in a way, simpler than in our previous work as many
reaction variables have been eliminated out of the search. After
removing the need for quantum effects and for thermodynamic
equilibrium, it is clear that isoscaling should, then, exist in
systems with little more than protons and neutrons without specific
interactions or dynamics. The origin of isoscaling must then lie in
the sampling (\textit{i.e.} mode of fragmentation) of a conglomerate
of protons and neutrons and perhaps, since we are dealing with
finite systems, on its geometry and homogeneity.

This work aims at elucidating the origin of isoscaling by searching for this
effect in a system with the bare minimum number of ingredients, namely bond
percolation model. After presenting the model in the next section, results
of the percolation of hundreds of thousands of three-dimensional grids are
presented in section~\ref{num}, followed with a number of conclusions in
section~\ref{concl}.

\bigskip

\section{The percolation model}

\label{perco} In order to explore the behavior of isoscaling we use
a three dimensional bond percolation model. This model was first
applied in nuclear multifragmentation by Bauer \textit{et
al.}~\cite{bauer} and used by many
groups~\cite{campi,staufer,elliott,cardenas} ever since. In the
usual bond percolation model, a fragmenting nucleus is represented
by a three-dimensional cubic lattice, and individual
\textquotedblleft nucleons\textquotedblright\ by nodes on the
lattice. All nodes start with bonds to all nearest-neighbors, which
represent nucleon-nucleon interactions. These bonds are then
attempted to be broken statistically according to a probability $b$,
thus producing clusters of connected nodes which are interpreted as
fragments.

To this usual model, we add the ``isospin'' of the nodes as an extra
degree of freedom. With this method, lattices with different ratios
of ``protons'' and ``neutrons'' can be constructed and ruptured,
producing cluster yields which can then be used to construct the
ratio $R_{21}$. Besides the usual assumptions of bond percolation,
the following three conditions are added in this analysis: $i)$ the
bond breaking probability, $b$, is spatially homogeneous and
identical for $pp$, $pn$, and $nn$ bonds, $ii)$ the probability of a
node for having isospin ``up'' ($p$) or ``down'' ($n$) is spatially
homogeneous, and $iii$) the number of protons and neutrons is fixed
from the beginning.

\subsection{Geometrical arguments leading to isoscaling}

\label{geom} 
\begin{figure}[tbph]
\setlength{\abovecaptionskip}{40pt} \centering
\includegraphics[width=10cm,angle=0]{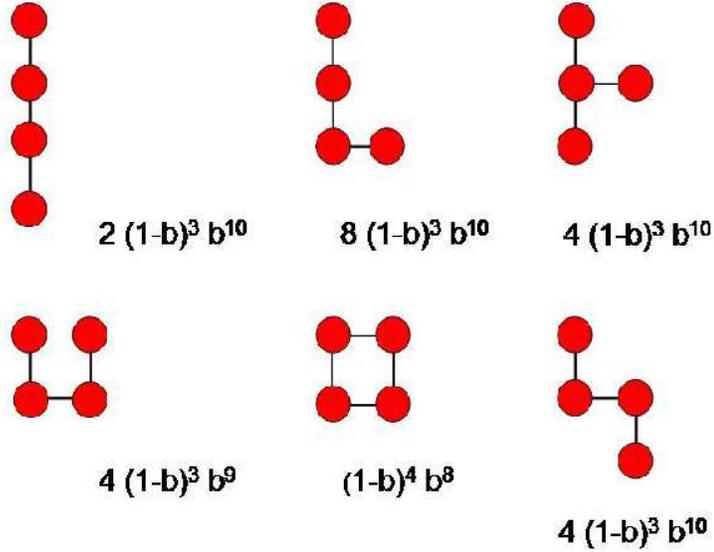}
\caption{Cluster structures in two dimensional percolation. In this figure
we show the 6 possible realizations of clusters of size 4 in a two
dimensional grid. Circles denote nodes, straight lines denote active links.
Together with each diagram we show the corresponding term in the sum $n_{4}$.
}
\label{fig1}
\end{figure}

In this model, fragments are obtained when bonds are broken
with a given probability $b$. If we consider the infinite size limit is
customary to express the number of fragments as the number of fragments per
node
\begin{equation}
\underset{L\rightarrow \infty }{\lim }\frac{N_{A}}{L^{3}}=n_{A} \ ,
\end{equation}
where $L^{3}$ is the size of the lattice measured in nodes, $N_{A}$
is the number of fragments of size $A$, and $n_{A}$ the number of
fragments of size $A$ per node. This last quantity can be written in
the following way
\begin{equation}
n_{A}=\underset{a,t}{\sum }g_{Aat}(1-b)^{a}b^{t} \ ,
\end{equation}
where $t$ stands for the perimeter of the cluster (number of bonds
to be broken in order to isolate the cluster composed by $A$ nodes),
and $a$ is the number of bonds linking the $A$ nodes, $g_{Aat}$ is
the number of cluster configurations with size $A,$ perimeter $t$
and $a$ activated bonds. As an illustration, figure~\ref{fig1} shows
the different terms appearing in the expression for $n_4$ in the two
dimensional case.  The resulting expression for this term is
\begin{equation}
n_{4}=14(i-b)^{3}b^{10}+4(1-b)^{3}b^{9}+(1-b)^{4}b^{8}
\end{equation}

To include the isospin degree of freedom, a given node belonging to a
cluster of size $A$ will be considered to be a proton with probability $p$.
Then
\begin{equation}
n_{A}=\left[ \underset{t}{\sum }g_{At}(1-b)^{a}b^{t}\right] \left[ \underset{%
Z=0,A}{\sum }\alpha _{Z}p^{Z}(1-p)^{(A-Z)}\right]
\end{equation}
with $\alpha _{Z}$ being the number of ways of building a cluster
with $Z$ protons and $N=A-Z$ neutrons: $\alpha _{Z}={A!}/{N!Z!}$ The
number of fragments per node with $A$ particles and $Z$ protons is
then defined as:
\begin{equation}
n_{A,Z} = n_{A}[\alpha _{Z}p^{Z}q^{(A-Z)}]=n_{A}\alpha
_{Z}p^{Z}q^{N} \ .
\end{equation}

Focusing now on the isoscaling problem, the quantity of interest is the
quotient:
\begin{equation}
R_{21}=\frac{Y_{2}(A,Z)}{Y_1(A,Z)}
\end{equation}
With $Y_{2}$ representing the yield for the reaction involving the
neutron rich nuclei.

For the percolation model, $R_{21}$ takes the form
\begin{equation}
R_{21}=\frac{[n_{A,Z}]_{2}}{[n_{A,Z}]_{1}}=\frac{p_{2}^{Z}q_{2}^{N}}{
p_{1}^{Z}q_{1}^{N}}=\left( \frac{p_{2}}{p_{1}}\right) ^{Z}\left(
\frac{q_{2} }{q_{1}}\right) ^{N}\ ,  \label{R12p}
\end{equation}
in which all the geometrical and combinatorial terms cancel out and only the
part related to the occupancy probabilities remains. In this way
\begin{equation}
R_{21}=\exp \left\{ \ln \left[ \left( \frac{p_{2}}{p_{1}}\right)
^{Z}\left( \frac{q_{2}}{q_{1}}\right) ^{N}\right] \right\} =\exp
\left[ N\ln \left( \frac{q_{2}}{q_{1}}\right) +Z\ln \left(
\frac{p_{2}}{p_{1}}\right) \right] =\exp \left( \alpha N+\beta
Z\right) \ .
\end{equation}
Which is the standard expression of the isoscaling coefficient, but now with
a clear microscopic interpretation for the isoscaling parameters $\alpha $
and $\beta $.

For finite lattices, it is convenient to study $R_{21}$ calculated in terms
of the number of fragments $N_{A,Z}$ instead of the above derivation in
terms of the number of fragments per node, $n_{A,Z}$. This can be achieved
by using the above derived expressions for the constants $\alpha $ and $%
\beta $. Equation~(\ref{R12p}) can be rewritten as
\begin{equation}
R_{21}=\frac{\left[ N_{A,Z}\right] _{2}}{\left[ N_{A,Z}\right] _{1}}\approx
\frac{A_{2}[n_{A,Z}]_{2}}{A_{1}[n_{A,Z}]_{1}}=\frac{A_{2}}{Z}\frac{Z}{A_{1}}%
\left( \frac{p_{2}}{p_{1}}\right) ^{Z}\left( \frac{q_{2}}{q_{1}}\right)
^{N}=\left( \frac{p_{2}}{p_{1}}\right) ^{Z-1}\left( \frac{q_{2}}{q_{1}}%
\right) ^{N}
\end{equation}%
where the probabilities have been approximated by $p_{i}\simeq Z/A_{i}$.
This immediately yields the usual expression:
\begin{equation}
R_{21}=C\exp \left( \alpha N+\beta Z\right)
\end{equation}%
with $\alpha =\ln (q_{2}/q_{1})$, $\beta =\ln (p_{2/}p_{1})$ and $C=\exp
[\ln (p_{1}/p_{2})]=p_{1}/p_{2}=A_{2}/A_{1}$

It should be kept in mind that the main assumptions in this derivation are
that the probability $p$ is homogenous and that the number of fragments per
node (per particle) can be approximated by the corresponding infinite size
limit. We now turn to a numerical verification of these predictions.

\section{Numerical experiments}

\label{num}

To explore the isoscaling phenomena in the framework of the above defined
percolation model, two different types of calculations were performed:
fixing the number of protons in the lattice, and using a fixed probability
to assign the isospin to the nodes.

\subsection{Fixed proton number}

In this case, the $R_{21}$ was obtained using the yields of two lattices,
one of size $6\times 6\times 6$, \textit{i.e.} with $A=216$ nodes, and a
second one of $7\times 7\times 7$ with $A=343$ nodes, both with the number
of protons fixed to $108$. These grids, had, then, $Z=N=108$, and
probabilities $p_{1}=0.5$ and $q_{1}=(1-p_{1})=0.5$ for the $6\times 6\times
6$ grid, and $Z=108$, $N=235$, $p_{2}=0.315$ and $q_{2}=(1-p_{2})=0.685$,
for the $7\times 7\times 7$ grid.

For these probabilities, the isoscaling coefficients are $\alpha =\ln
(0.6852/0.5)=\allowbreak 0.315\,$ and $\beta =\ln (0.3148/0.5)=\allowbreak
-0.463$, and the constant $C=p_{1}/p_{2}=0.5/0.3148=\allowbreak
1.\,\allowbreak 588\,3$. The results of $200,000$ percolations of each of
these grids are displayed in panel $a)$ of figure~\ref{fig2}.

\subsection{Fixed $p$ and $n$ occupancy probability}

In this second case, the two isotopically different grids were constructed
with the same sizes for both, but with two different protons and neutrons
occupation probabilities. Several cases were studied.

\begin{figure}[tbph]
\setlength{\abovecaptionskip}{40pt} \centering
\includegraphics[width=11cm,angle=270,clip=]{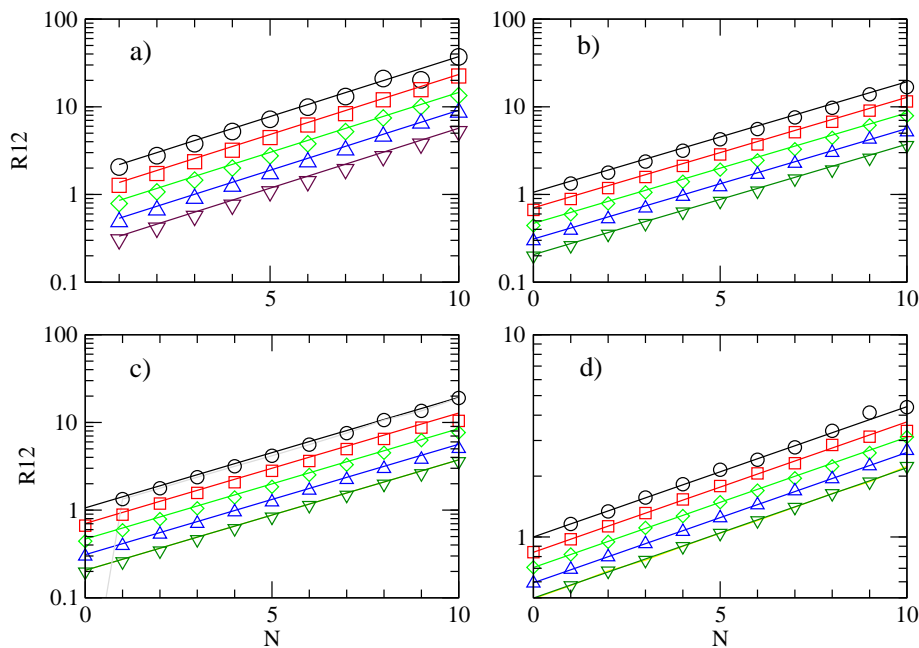}
\caption{ Isoscaling in percolation. Comparison of theoretical predictions
for isoscaling in the nuclear percolation model and the corresponding
results from numerical simulations. Lines denote theoretical predictions,
symbols denote numerical simulations}
\label{fig2}
\end{figure}

First, $6\times 6\times 6$ grids were constructed using $p_{1}=0.5$ and $%
q_{1}=(1-p_{1})=0.5$ which yields a total number of protons (in average) of $%
\left\langle Z\right\rangle =108$ and $\left\langle N\right\rangle =108$,
and $p_{2}=0.33$ and $q_{2}=(1-p_{2})=0.67$ which yields a total number of
protons (in average) of $\left\langle Z\right\rangle =$ $72$ and $%
\left\langle N\right\rangle =$ $144$. In this case the coefficients are : $%
\alpha =\ln (0.67/0.5)=\allowbreak 0.292\,67,$ and $\beta =\ln
(0.33/0.5)=\allowbreak -0.415\,52$. The results obtained from this numerical
exercise are displayed in panel b) of figure~\ref{fig2}; the fact that
isoscaling is well reproduced by this purely geometrical model is obvious
from this figure.

Next, to investigate size effects, a similar case was constructed using a
smaller lattice of size $5\times5\times5$ with the same probabilities as in
the previous case. The corresponding results are displayed in panel c) of
figure~\ref{fig2}; again the property of isoscaling is apparent.

Finally, to investigate the effect of different occupation probabilities on $%
R_{21}$, one more case was studied. In this case, a lattice of
$6\times 6\times 6$ was populated with protons and neutrons
according to the probabilities: $p_{1}=0.5$ and
$q_{1}=(1-p_{1})=0.5$. For the neutron rich partner the
probabilities used were $p_{2}=0.42$ and $q_{2}=(1-p_{2})=0.58$
which yields a total number of protons (on average) of $\left\langle
Z\right\rangle =$ $91$ and $\left\langle N\right\rangle =$ $125$. In
this case the coefficients are $\alpha =\ln (0.58/0.5)=\allowbreak
0.148\,42,$ and $\beta =\ln (0.42/0.5)=\allowbreak -0.174\,35$. The
corresponding results are shown in panel d) of figure~\ref{fig2}.

For each of these cases $200,000$ configurations were generated. The
bond breaking probability was chosen as $b=0.31$ in order to get
good statistics in an ample range of masses. [Attempts with
different values of the breaking probability $b$ demonstrated that
the ratio $R_{21}$ is independent of $b$.] The figure also shows a
comparison of the numerical results to the theoretical predictions
of section~\ref{num}, lines denote theoretical predictions whereas
symbols denote numerical simulations. The agreement between the
theoretical predictions and the results of the simulations is
remarkable.

\section{Conclusions}

\label{concl}

We have studied the isoscaling phenomenon in the frame of
percolation model. We have derived exact analytic expressions for
the infinite case and approximate ones for the finite case. We have
performed numerical simulations for not too big systems ($125,$
$216$ and $343$ \textquotedblleft particles\textquotedblright ) with
different relative populations of N:Z. The excellent agreement
between numerical simulations and theory indicate that isoscaling
emerges from the simple assumption of fair sampling with homogeneous
probabilities. On the other hand, this property can be seen as a
minimum information approach, \textit{i.e.} all configurations are
equiprobable, as such this analysis can be interpreted in the frame
of a maximum entropy approach. This indicates that the information
about effects due, for example, to the asymmetry term in the
equation of state, is in the absolute values of the parameters
$\alpha $ and $\beta$, and not in the isoscaling property itself.

\begin{acknowledgments}
C.O.D. acknowledges the support of Universidad de Buenos Aires
through grant x139, CONICET through grant 2041, and the hospitality
of the University of Texas at El Paso.
\end{acknowledgments}

\end{document}